\documentclass[%
 reprint,
 amsmath,amssymb,
 aps,
 pra,
]{revtex4-2}

\usepackage{graphicx}
\usepackage{dcolumn}
\usepackage{bm}

\usepackage{bm}
\usepackage{amsfonts}
\usepackage{mathrsfs}
\usepackage[colorlinks, citecolor=red]{hyperref}

\begin{document}

\title{High dimensional entanglement between a photon and a multiplexed \\ atomic quantum memory}

\author{C. Li$^{}$}
\affiliation{Center for Quantum Information, IIIS, Tsinghua University, Beijing 100084, PR China}

\author{Y.-K. Wu$^{}$}
\affiliation{Center for Quantum Information, IIIS, Tsinghua University, Beijing 100084, PR China}

\author{W. Chang$^{}$}
\affiliation{Center for Quantum Information, IIIS, Tsinghua University, Beijing 100084, PR China}

\author{S. Zhang$^{}$}
\affiliation{Center for Quantum Information, IIIS, Tsinghua University, Beijing 100084, PR China}

\author{Y.-F. Pu$^{}$}
 \altaffiliation[Present address: Institute for Experimental Physics, University of Innsbruck, A-6020 Innsbruck, Austria. ]{}

\author{ N. Jiang$^{}$}
\altaffiliation[Present address: Department of Physics, Beijing Normal University, Beijing 100875, China]{}

\author{L.-M. Duan$^{\footnotemark[3]}$}
\email{Correspondence should be addressed to lmduan@tsinghua.edu.cn}
\affiliation{Center for Quantum Information, IIIS, Tsinghua University, Beijing 100084,
PR China}

\begin{abstract}
Multiplexed quantum memories and high-dimensional entanglement can improve the performance of quantum repeaters by promoting the entanglement generation rate and the quantum communication channel capacity. Here, we experimentally generate a high-dimensional entangled state between a photon and a collective spin wave excitation stored in the multiplexed atomic quantum memory. We verify the entanglement dimension by the quantum witness and the entanglement of formation. Then we use the high-dimensional entangled state to test the violation of the Bell-type inequality. Our work provides an effective method to generate multidimensional entanglement between the flying photonic pulses and the atomic quantum interface.
\end{abstract}

\maketitle


\section{Introduction}

Long distance quantum communication requires quantum entanglement distributed over two end nodes of a quantum communication channel \cite{ekert1991quantum, kimble2008quantum, wehner2018quantum}. Due to the optical absorption and other noise in the channel, the error of direct communication increases exponentially with the distance, thus reduces the key rates in quantum key distribution. To overcome this problem, the quantum repeater protocol has been proposed, where a series of entanglement generation and swapping operations are performed to extend the entanglement to farther and farther nodes with only polynomial cost \cite{briegel1998quantum}. The practical utilization of a quantum repeater requires quantum memories \cite{duan2001long, bussieres2013prospective}. Pioneering works have been demonstrated toward the implementation of a quantum repeater with atomic quantum memory. For example, photonic qubits have been stored as collective spin wave excitations in the atomic ensemble \cite{hsiao2018highly, wang2019efficient, usmani2010mapping}; entanglement between the memory and transmitting photons has also been realized \cite{matsukevich2005entanglement, farrera2018entanglement, PhysRevX.9.041033}.

Several methods have been proposed to further improve the quantum repeater protocol. One is to use multiplexed quantum memories, which significantly reduce the required time to establish entanglement in the quantum communication channel \cite{collins2007multiplexed, lan2009multiplexed, parniak2017wavevector}. Another possibility is to explore high-dimensional entanglement in the quantum network \cite{bechmann2000quantum, cerf2002security}, which increases the capacity of the communication channel and thus enhances the quantum communication efficiency \cite{simon2007quantum}. High-dimensional entanglement also has plenty of applications beyond quantum communication, such as quantum teleportation with high capacity \cite{steinlechner2017distribution, hu2019experimental, PhysRevLett.123.070505}, quantum distillation \cite{li2014entanglement, kwiat2001experimental} and robust Bell tests \cite{dada2011experimental, wang2018multidimensional}. Recently, many efforts have been devoted to creating high-dimensional entanglement sources in different systems, like rare-earth-doped crystals \cite{kutluer2017solid, ikuta2018four, martin2017quantifying, tiranov2017quantification, edgar2012imaging, schneeloch2019quantifying, kues2017chip}, integrated devices \cite{wang2018multidimensional, kues2017chip} and atomic systems \cite{wen2019multiplexed, ding2016high, pan2019orbital}.
Moreover, the photonic qudits possessing the high-dimensional entanglement have been stored in quantum memory elements, based on atomic ensembles \cite{parigi2015storage, zhang2016experimental} or rare-earth-doped crystals \cite{PhysRevLett.123.080502, usmani2010mapping,kutluer2017solid}.

Despite its importance, the verification of a high-dimensional entangled state and the certification of the entanglement dimension are sophisticated tasks for the experiments. The standard method to estimate the entanglement fidelity is to reconstruct the full quantum state, for example, through quantum state tomography \cite{PhysRevA.64.052312, thew2002qudit}. It works well for low-dimensional entangled states \cite{ikuta2018four}; but the measurement costs increase significantly with the system dimension. Besides being time consuming, it also requires the setup to be stable during the whole measurements. Furthermore, some measurement settings may not be available for a given experimental platform \cite{tiranov2017quantification, schneeloch2019quantifying}. To overcome these difficulties, several methods are proposed and experimentally demonstrated to efficiently characterize the high-dimensional entanglement with sparse data, such as entanglement witness \cite{dada2011experimental, tiranov2017quantification, schneeloch2019quantifying, krenn2014generation} and compressed sensing \cite{gross2010quantum, riofrio2017experimental}. 

In this paper, we experimentally demonstrate an alternative and effective method to generate high-dimensional entanglement between a flying photon pulse and a spin wave stored in an atomic quantum memory based on the use of spatial multiplexing. Through excitation of a one-dimensional (1D) array of $10$ atomic memory cells \cite{pu2017experimental, jiang2019experimental}, we generate high-dimensional entanglement carried by different spatial modes of the photon and the atoms. These different modes are brought together for interference through an acoustic-optical deflector (AOD) to confirm the multi-dimensional entanglement. The high-dimensional entanglement is verified through different kinds of entanglement witnesses, and we confirm that at least eight-dimensional entanglement is achieved experimentally \cite{krenn2014generation}. Entanglement of formation is also measured to confirm a lower bound of 4-dimensional entanglement \cite{tiranov2017quantification}. Finally, as an application, we demonstrate the violation of high-dimensional Bell-type inequalities using the stored entangled state \cite{collins2002bell}.

\begin{figure*}[htb]
  \centering
  \includegraphics[width=18cm]{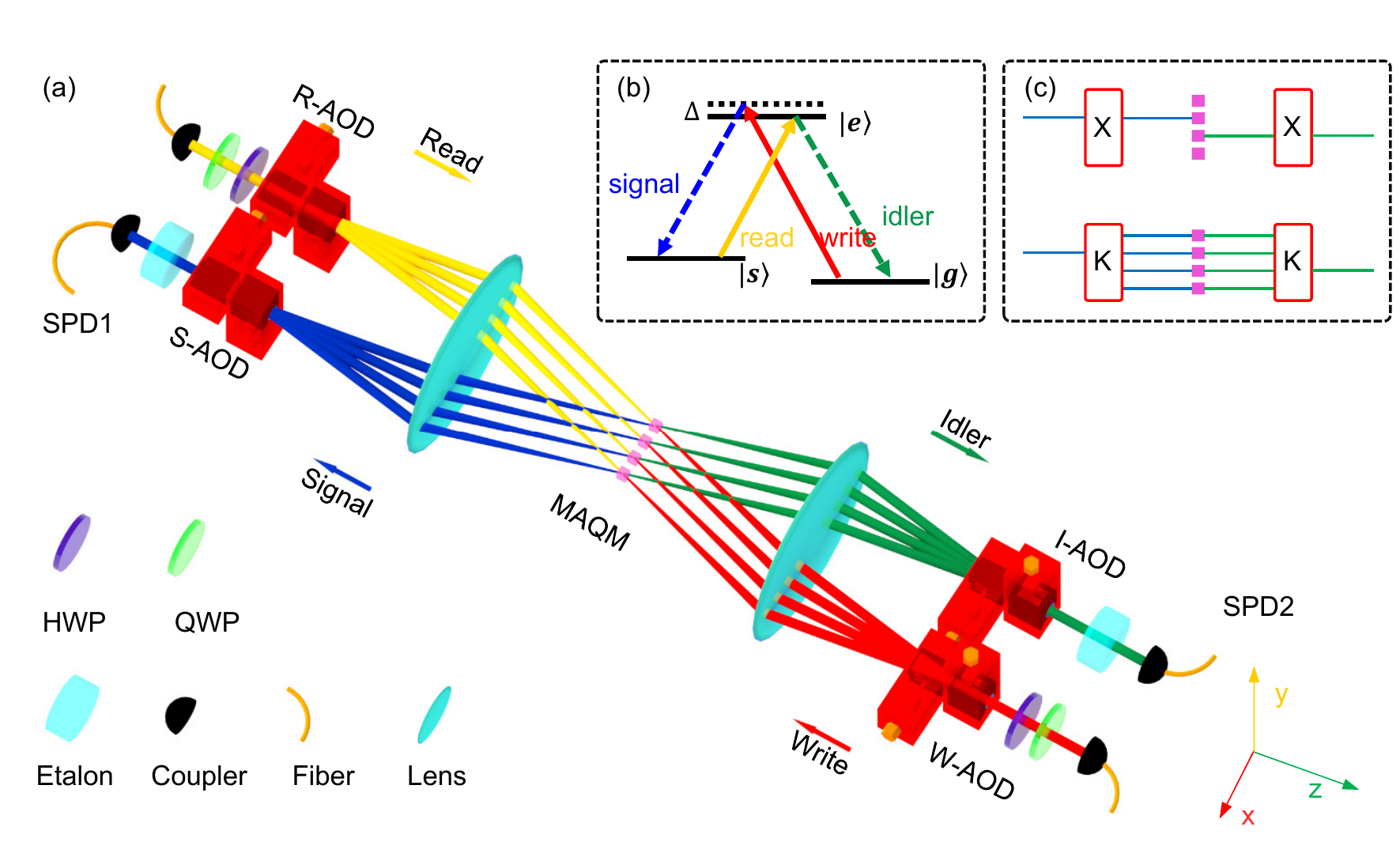}\\
\caption{\textbf{Generation of high dimensional entanglement between a photon and a multiplexed atomic quantum memory (MAQM).} \textbf{(a)} Setup. HWP represents half-wave plate, QWP for quarter-wave plate, and SPD for single photon detector. The two programmable acousto-optic deflectors (AOD) (W-AOD/R-AOD or S-AOD/I-AOD), two lenses and the atomic ensemble in the middle under 4f configuration form the multiplexing/demultiplexing optical circuits. The write pulse is split into ten paths in the $x$ direction (4 paths shown here for clarity). The entanglement between a signal photon and a spin wave excitation in the memory array is generated by the DLCZ protocol. To verify the entanglement, the spin wave is retrieved into an idler photon. The ten optical modes of the signal/idler photon are further combined in the S-AOD/I-AOD; measurements in different bases can be performed by adjusting the amplitudes and phases of the RF signals on the AODs. \textbf{(b)} Energy diagram $|g\rangle\equiv|5S_{1/2}, F=1\rangle$, $|s\rangle\equiv|5S_{1/2}, F=2\rangle$, $|e\rangle\equiv|5P_{1/2}, F=2\rangle$. The write beam is blued-detuned by $\Delta=16\,$MHz from resonance at the central memory cell. \textbf{(c)} Basis choice for AODs. the blue (green) lines are the signal (idler) modes; the pink squares represent the individual memory cells, and the red rectangles are the S-AOD/I-AOD. The AODs can be set to measure the X-basis (upper row), or the K-basis (lower row). More details can be found in Appendix~B.}
\end{figure*}

\begin{figure*}[tbp]
  \centering
  \includegraphics[width=18cm]{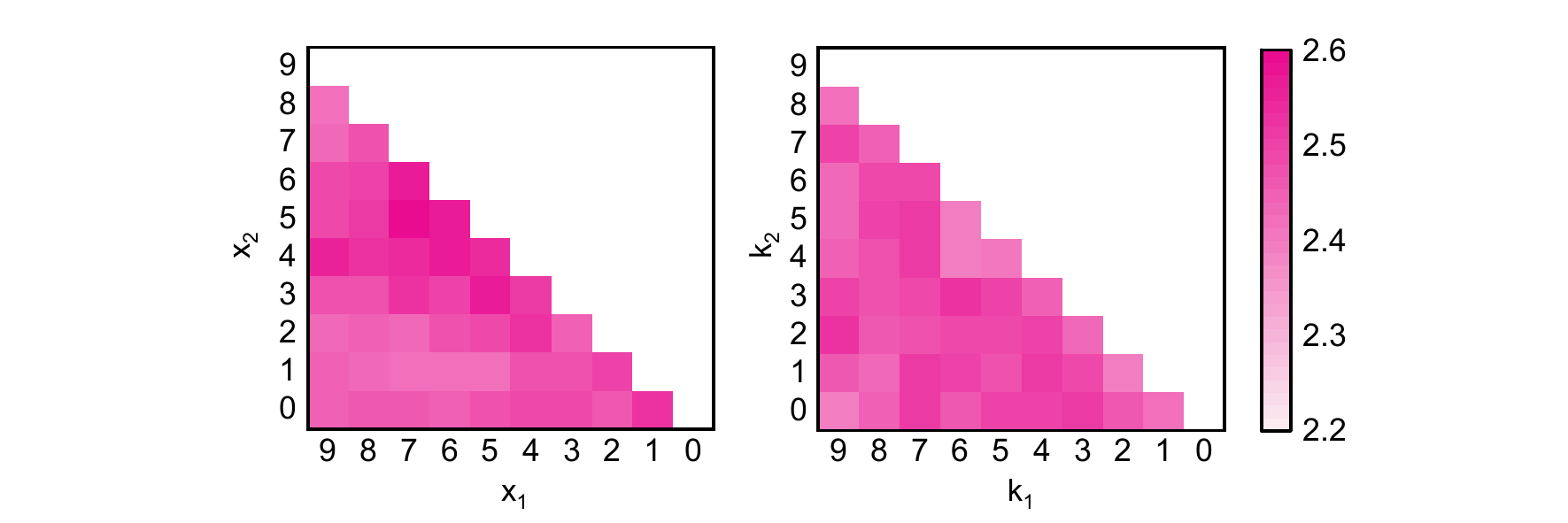}\\
\caption{\textbf{Measurement of entanglement witness.} The sum of visibility $V_x$, $V_y$ and $V_z$ in X-space and K-space are shown respectively. Each pixel in the plot corresponds to a two-dimensional subspace spanned by $X=x_1$ and $X=x_2$ ($K=k_1$ and $K=k_2$) for both the signal and the idler photons. Every mode pair requires 12 measurement settings to obtain the visibility, and the ideal sum of visibility is 3 for each pair. }
\end{figure*}

\section{Experimental setup}
Our experiment is illustrated in Fig.~1 schematically. A cold $^{87}$Rb ensemble is loaded into a 3D MOT with about 2-3 billion atoms. After a $10\,$ms compression and a $7\,$ms polarization gradient cooling (PGC) processes, the temperature of the ensemble is reduced to $25\,\mu$K. Then we further prepare the atoms to the ground state $|g\rangle\equiv|5S_{1/2}, F=1\rangle$ through optical pumping. The experiment starts with a $100\,$ns write pulse, which is $16\,$MHz blue-detuned to the D1 transition $|g\rangle\rightarrow|e\rangle\equiv|5P_{1/2}, F=2\rangle$. Upon the detection of a signal photon from spontaneous Raman scattering, we know that an atom has been scattered into the storage state $|s\rangle\equiv|5S_{1/2}, F=2\rangle$, and that a spin wave has been generated in the atomic ensemble. Then it can be retrieved back into an idler photon by a $500\,$ns strong read pulse, resonant to the $|s\rangle\rightarrow|e\rangle$ transition. The interval time between the write and the read pulse, $7.9\,\mu$s, is the same as the Larmor period of the ensemble. In this way, we can achieve the highest retrieval efficiency of the spin wave excitation \cite{li2019quantum}.
On the other hand, if there is no signal photon detected following the write pulse, a strong clean pulse identical to the read pulse will be applied to bring the atoms back to the ground state $|g\rangle$, and the experimental cycle will be repeated.

The multiplexing/demultiplexing optical circuits consists of the acousto-optic deflectors (AODs) and lenses under 4f configuration to address different regions of the cold $^{87}$Rb ensemble. Each region serves as an individually addressable memory cell with low cross talk errors \cite{pu2017experimental}. We use the Duan-Lukin-Cirac-Zoller (DLCZ) protocol to generate the high dimensional entanglement between the signal photon and the spin wave in the memory array \cite{duan2001long}. In this work, we divide the weak write pulse into 10 spatial modes (4 shown in Fig.~1(a) for clarity). Since the signal photon and the spin wave come from the same spatial mode, the entangled state can be written as:
\begin{equation}
|\Psi\rangle = \sum_{i=0}^{9}C_i|i\rangle_s|i\rangle_a,
\end{equation}
where $|i\rangle_s$ $(|i\rangle_a)$ refers to the signal photon (spin wave) in the mode $i$, and the coefficients $C_i$ can be controlled by adjusting the amplitudes and phases of RF tones in the W-AOD. Ideally we want $C_i=1/\sqrt{10}$ ($i=0,\,1,\,\cdots,\,9$) for the maximally entangled state.

To verify the high-dimensional entanglement, we first retrieve the spin wave excitation into an idler photon by a strong read pulse. The signal modes and the idler modes are combined by the S-AOD and I-AOD respectively for different measurement bases. Two types of bases will be used: the spatial basis $\hat{X}=|x\rangle\langle x|$ ($x=0,\,1,\,\cdots,\,9$), and the momentum basis $\hat{K}=|k\rangle\langle k|$, where $|k\rangle = \sum_{x=0}^{9} \exp (2\pi ixk/10)|x\rangle/\sqrt{10}$ ($k=0,\,1,\,\cdots,\,9$).

\section{Entanglement witness}

Quantum state tomography can reconstruct the full density matrix, but it is time-consuming for high-dimensional entangled states. Here, to efficiently verify the entanglement, we use the entanglement witness method \cite{ding2016high, krenn2014generation}. The entanglement witness we use here was originally designed for photons carrying orbital angular momentum (OAM). Three mutually unbiased bases (MUBs) are measured for every 2-dimensional subspace; following Ref.~\cite{ding2016high}, we call them diagonal/anti-diagonal ($\sigma_x$), left/right ($\sigma_y$), and horizontal/vertical ($\sigma_z$) bases. The visibility is defined as the correlation between the signal and the idler photons $V_i=|\sigma_i^{(s)} \otimes \sigma_i^{(i)}|$ ($i=x, y, z$). We measure the visibility for all the two-dimensional subspaces in the X (K) space spanned by $X=x_1$ and $X=x_2$ ($K=k_1$ and $K=k_2$), as shown in Fig.~2. If the system has at most $d$-dimensional entanglement, it can be shown that the sum of visibilities for all the mode pairs $W$ satisfies \cite{ding2016high}:
\begin{equation}
W \le f(d) \equiv 3D(D - 1) / 2 - D(D - d),
\end{equation}
where $D=10$ is the total number of modes. When the measured $W$ exceeds this bound, we can conclude that the entanglement is at least $(d+1)$-dimensional. Here we measure $W_X=111.6\pm0.8$ for the X-space, and $W_K=111.1\pm0.8$ for the K-space. Both of them are larger than $f(d)=105$ for $d=7$, which verifies that at least an 8-dimensional entangled state is generated. The major obstruction to reach higher visibility is the accidental coincidence counts \cite{de2006direct, jing2019entanglement}. It reduces the signal-to-noise ratio and hence the visibility of all the two-dimensional subspaces. If we subtract the accidental coincidence from the data, as described in Appendix~C, the calculated total visibilities are $W_{X}^{\prime}=126.5\pm 1.0$ and $W_{K}^{\prime}=125.5\pm 0.9$, which correspond to at least 10- and 9- dimensional entanglement, respectively.

\begin{figure}[ptb]
  \centering
  \includegraphics[width=8.6cm]{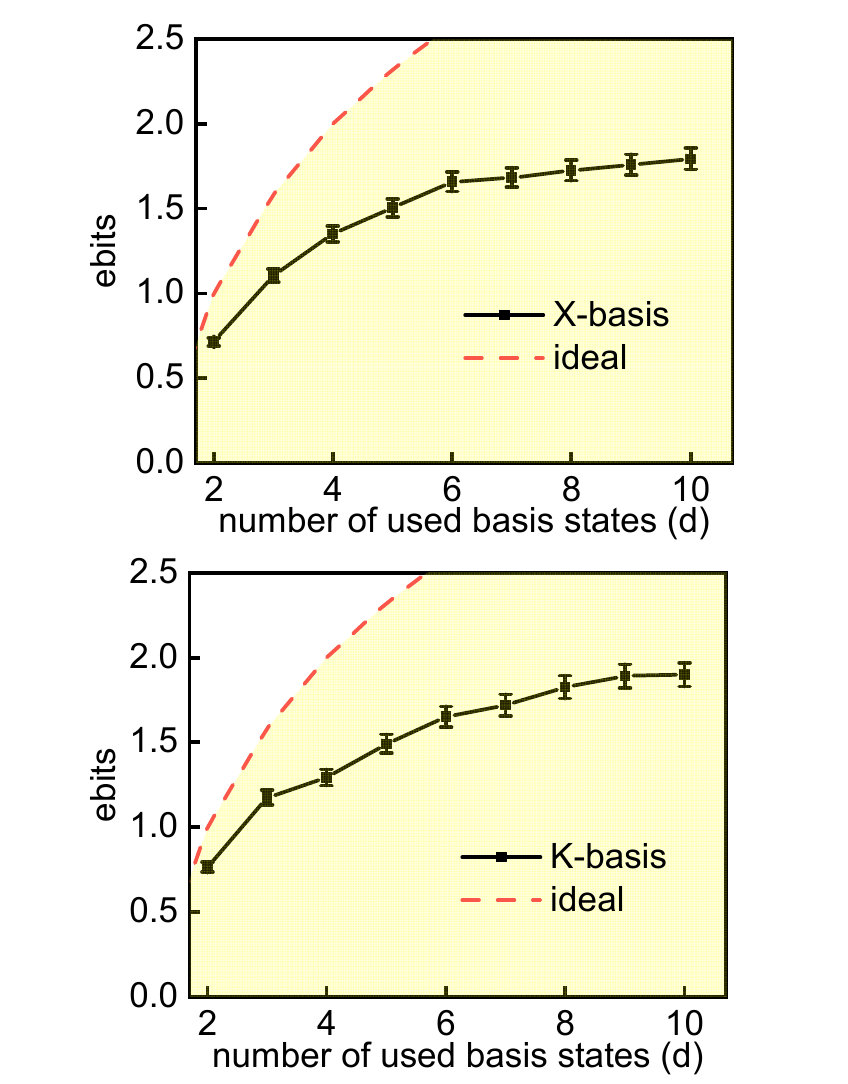}\\
\caption{\textbf{Entanglement of formation.} The entanglement of formation is measured for both X-basis and K-basis after subtracting accidental coincidence. The maximal values are 1.793 ebits for the $X$ basis and 1.90 ebits for the $K$ basis. The red dashed line represents the results for an ideal maximally entangled state with $\mathrm{log}_2 d$ ebits.  }
\end{figure}

\section{entanglement of formation}
We also bound the entanglement of formation $E_F$ to verify the entangled state. The entanglement of formation gives the minimal number of maximally entangled qubit-qubit states (ebits) that is required to get one copy of the desired entangled state through local operations and classical communication (LOCC) \cite{martin2017quantifying, tiranov2017quantification}. A lower bound of $E_F$ is given by \cite{tiranov2017quantification}
\begin{equation}
E_F\ge- \log_2(1-B^2/2),
\end{equation}
where $B$ is defined as
\begin{align}
B=\frac{2}{\sqrt{|C|}} \sum_{(j,k) \in C\atop j< k} &\Big( |\langle j,j|\rho|k,k\rangle|\nonumber\\
&-\sqrt{\langle j,k|\rho|j,k\rangle\langle k,j|\rho|k,j\rangle} \Big).
\end{align}
In the above equation, $\rho$ is the density matrix of the entangled state; $j$ $(k)$ indicates the mode $j$ $(k)$ of the signal and the idler photons; $C$ is a set of mode pairs $(j,\,k)$ and $|C|$ denotes the number of pairs in the set. To lower bound $ E_F$, we need to measure the $\langle j,k|\rho|j,k\rangle$ and $\langle j,j|\rho|k,k\rangle$ terms, which can be obtained in a similar way as the visibility in the entanglement witness experiment. Note that in Ref.~\cite{tiranov2017quantification} the $\langle j,j|\rho|k,k\rangle$ terms for $|j-k|>1$ are bounded by the nearest neighbor terms, and the bounds become looser for farther away pairs; hence the bound on the entanglement of formation ceases to improve for these pairs. In comparison, here we are able to measure all these terms directly; since the entangled state we prepare has high fidelity (even though we are not able to measure the fidelity by quantum state tomography), it will be beneficial to include all the mode pairs in Eq.~(8). In Fig.~3 we show the increase of the bound on the entanglement of formation as we consider more and more X or K modes. In this way, we get the tightest bound on the entanglement of formation as $1.79\pm0.06$ and $1.90\pm0.07$ for the X-space and K-space measurements respectively (after subtraction of accidental coincidence; details can be found in Appendix~C). This result verifies genuine 4-dimensional entanglement.

\section{Bell-type inequality}
Finally, we use our entangled state to study the violation of high-dimensional Bell-type inequalities. The original Bell inequality focuses on a qubit-qubit entangled state \cite{bell1964physics}. It has been generalized to higher-dimensional entangled systems with the advantage of stronger resistance to experimental noise \cite{kaszlikowski2000violations, collins2002bell}. Consider a bipartite quantum system of dimension $d\times d$. If the correlation can be described by local hidden variable theories, the Bell-type parameter $S_d$ will satisfy the CGLMP inequalities \cite{dada2011experimental, collins2002bell}:
\begin{widetext}
\begin{align}
S_d\equiv&\sum_{l=0}^{[d/2]-1}\left(1-\frac{2l}{d-1}\right)
\Big\{[P(I_0=S_0+l)-P(I_0=S_0-l-1)]+[P(S_1=I_0+l)-P(S_1=I_0-l-1)]\nonumber\\
&\qquad\qquad\qquad\qquad +[P(S_0=I_1+l+1)-P(S_0=I_1-l)]+[P(I_1=S_1+l)-P(I_1=S_1-l-1)]\Big\}\nonumber\\
\le& 2.
\end{align}
\end{widetext}
Here two possible detector settings can be used for the signal and the idler photons respectively, and for each detector setting there are $d$ possible measurement outcomes: $S_s,\,I_i=0,\,\cdots,\,d-1$ ($s,\,i=0,\,1$ represent the measurement settings). $P(I_i=S_s)$ is the probability that the signal photon and the idler photon outcomes are the same:
\begin{equation}
P(S_s=I_i)=\sum_{k=0}^{d-1}P(S_s=k,I_i=k).
\end{equation}
Similarly
\begin{equation}
P(S_s=I_i+l)=\sum_{k=0}^{d-1}P(S_s=k,I_i=(k-l)\text{ mod }d).
\end{equation}

The measurement bases for the signal and the idler photons are
\begin{equation}
|k\rangle_{S,s}=\frac{1}{\sqrt{d}}\sum_{x=0}^{d-1}\mathrm{exp}[2\pi ix(k+\Theta_s)/d]|x\rangle_S,
\end{equation}
\begin{equation}
|l\rangle_{I,i}=\frac{1}{\sqrt{d}}\sum_{x=0}^{d-1}\mathrm{exp}[2\pi ix(-l+\Phi_i)/d]|x\rangle_I,
\end{equation}
where $k,\,l=0,\,\cdots,\,d-1$ and $\Theta_0=0$, $\Theta_1=0.5$, $\Phi_0=0.25$, $\Phi_1=-0.25$.

\begin{figure}[ptb]
  \centering
  \includegraphics[width=8.6cm]{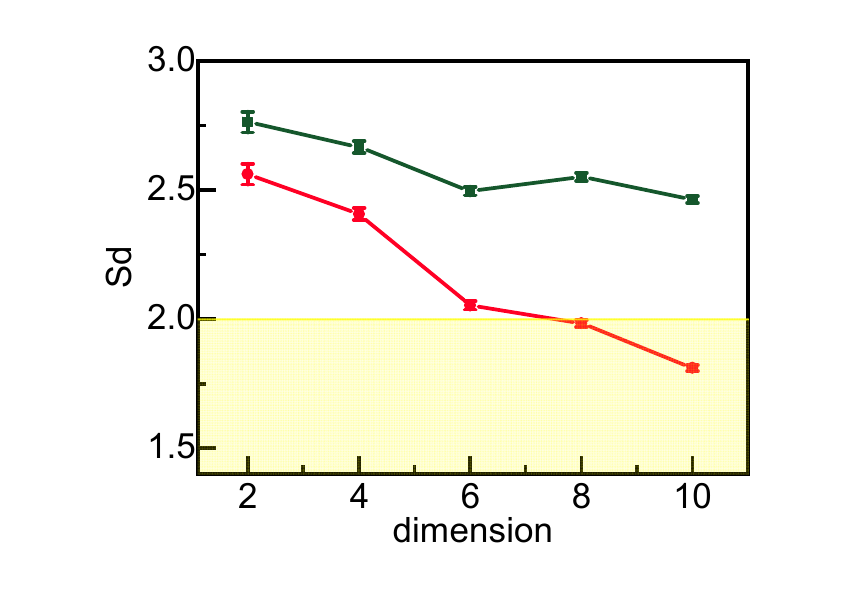}\\
\caption{\textbf{Violation of the CGLMP inequalities.} The error bars of the Bell parameter $S_d$ are estimated assuming a Poisson distribution of the photon counting. The violation of the Bell-type inequalities is observed for up to $d=6$ (red line), and up to $d=10$ after subtracting the accidental coincidence (green line).}
\end{figure}

Figure~4 shows the measured $S_d$ as a function of the dimension $d$, comparing with the classical bound. The violation of the CGLMP inequality remains up to $d=6$ (red line), and it can be extended to $d=10$ (green line) when the accidental coincidence is subtracted. This reveals the quantum nature of the correlation between the signal photon and the idler photon (thus the spin wave excitation in the atomic ensemble).

\section{Discussion and Conclusion}

The measured entanglement dimension is mainly affected by two factors: the accidental coincidence of photon counting on the signal and the idler detectors (details in Appendix~C), and the difference in the amplitude and phase settings on different AODs (details in Appendix~B and Appendix~D). There are also high order excitations and background noises, which are not the dominant errors in our experiment. To improve the visibility, the cross correlation $g_2$ between the signal and the idler photons can be increased by reducing the generation rate of the signal photon, so long as the coincidence is kept much higher than the background noise. In principle, a 2D array of memory cells can be used to create a higher dimensional entangled state, but there the visibility measurement will be difficult for two memory cells that are neither in the same row nor in the same column. Although the visibility can still be bounded based on the methods of Ref.~\cite{tiranov2017quantification}, it is not tight enough to give larger number of ebits or witness parameter $W$.

To summarize, in this work, we generate an entangled state between a signal photon and a spin wave excitation in a 1D MAQM array with 10 spatial modes. Entanglement witness and entanglement of formation are used to verify the entanglement, and we confirm at least 8- and 4-dimensional entanglement respectively using these two methods. The Bell-type inequality is studied as an application, which in turn proves the existence of entanglement in the system. Our experiment is an important step toward quantum repeaters and quantum networks using multiplexed quantum memories and high-dimensional entanglement. If each memory cell can store a photon carrying other degrees of freedom \cite{zhang2016experimental}, we can combine the advantages of high dimensional entangled state and multiplexed quantum memory together and expect further improvement in the performance.

\begin{acknowledgments}
This work was supported by the Ministry of Education of China, Tsinghua University, and the National key Research and Development Program of China (2016YFA0301902). Y.K.W. acknowledges support from Shuimu Tsinghua Scholar Program and International Postdoctoral Exchange Fellowship Program (Talent-Introduction Program).
\end{acknowledgments}

\begin{figure*}[ht]
  \centering
  \includegraphics[width=18cm]{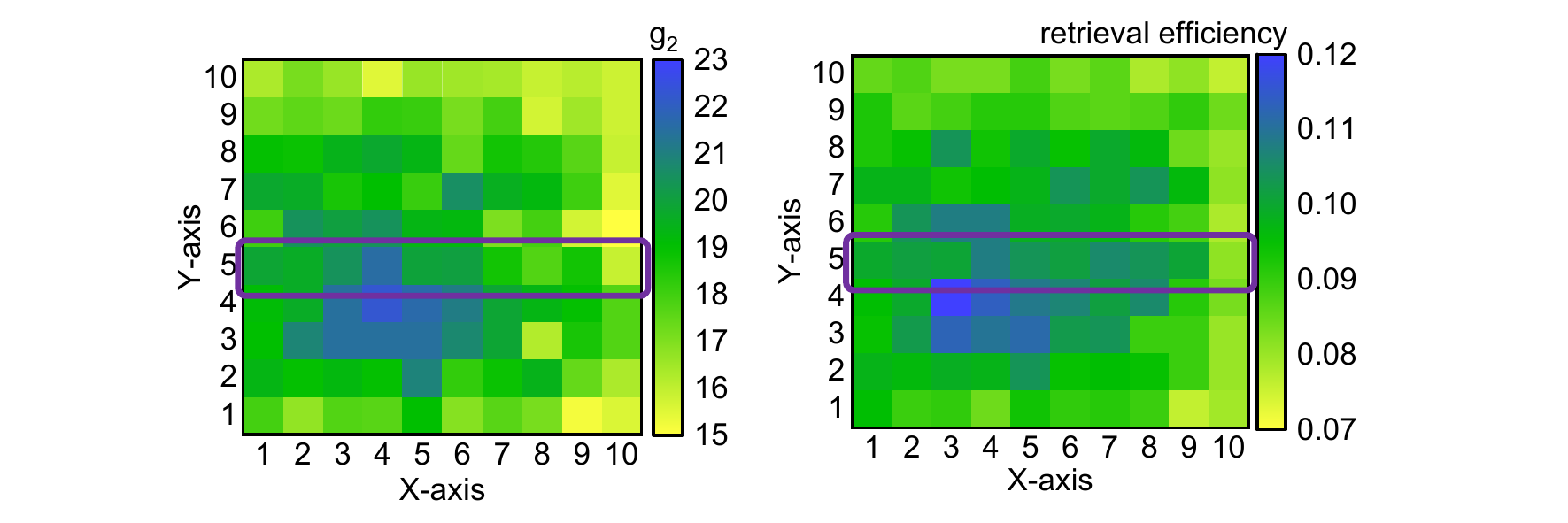}\\
\caption{\textbf{The $g_2$ correlation and retrieval efficiency of the MAQM.} The $g_2$ correlation and retrieval efficiency of the MAQM is measured before the experiment. The collecting probability of a signal photon is about $0.6\%$. We choose one $X$ row (specified by the purple box) to carry out the experiment. Its $g_2$ correlation and retrieval efficiency for each memory cell are both high enough and close to each other.}
\end{figure*}

\appendix


\section{The control of AODs}

The multiplexing/demultiplexing optical circuits are composed of a pair of 2D AODs and lenses under a 4f configuration. The AODs are controlled by the RF signals generated by arbitrary waveform generators (AWG). Before the experiment, we need to choose an optimized array with the highest and nearly uniform retrieval efficiency and optical depth for each memory cell. The cross correlation and retrieval efficiency for the 2D array are shown in Fig.~5 and the 1D subarray used for the experiment is specified in the purple box. The frequency interval between adjacent spatial modes in the $X$/$Y$ directions is $0.8\,$MHz on the AODs, and correspondingly the distance between the neighbor cells is $180\,\mu$m. The waists of the signal/idler and the write/read modes are $70\,\mu$m and $120\,\mu$m respectively, which guarantees a cross talk error well below $1\%$.

\section{AOD setting and measurement bases}
In the experiment, $d$ memory cells are addressed at the same time, therefore $d$ RF tones are input into the AODs. The nonlinearity of AODs will then produce two-tone third-order components \cite{hecht1977multifrequency}, which can be overcome by tuning the amplitude and phase of each RF tone \cite{antonov2007efficient, endres2016atom}. Once we have the optimized amplitude and phase settings for the $d$ RF components, a $K$-space basis state is created as the superposition of $d$ spatial modes:
\begin{equation}
|0\rangle_K\equiv\sum_{x=0}^{d-1}A_x e^{i\Phi_x}|x\rangle_X.
\end{equation}
Actually, the optimal phase setting is almost the same for the $d$-tone RF signals regardless of the interval frequency difference. The amplitude and phase settings should be optimized for different AODs to suppress the unwanted nonlinear effects and to unify the amplitude $A_x$ for each mode. After absorbing the phases into the definition of the spatial basis states, we can express this $K$-space basis state as
\begin{equation}
|0\rangle_K=\frac{1}{\sqrt{d}}\sum_{x=0}^{d-1}|x\rangle_X.
\end{equation}
Then the other $d-1$ $K$-space basis states can be obtained by adding proper phases on each RF tone as
\begin{equation}
|k\rangle_K=\frac{1}{\sqrt{d}}\sum_{x=0}^{d-1}\exp(2\pi ixk/d)|x\rangle_X
\end{equation}
for $k=1,\,\cdots,\,d-1$. With this method, bases used for all the AODs can be determined for arbitrary dimension $d=1,\,\cdots,\,10$. For W-AOD and R-AOD, we keep using the $|0\rangle_K$ ($d=10$) setup to generate the entangled state and to retrieve the spin wave.

\begin{figure*}[ptb]
  \centering
  \includegraphics[width=18cm]{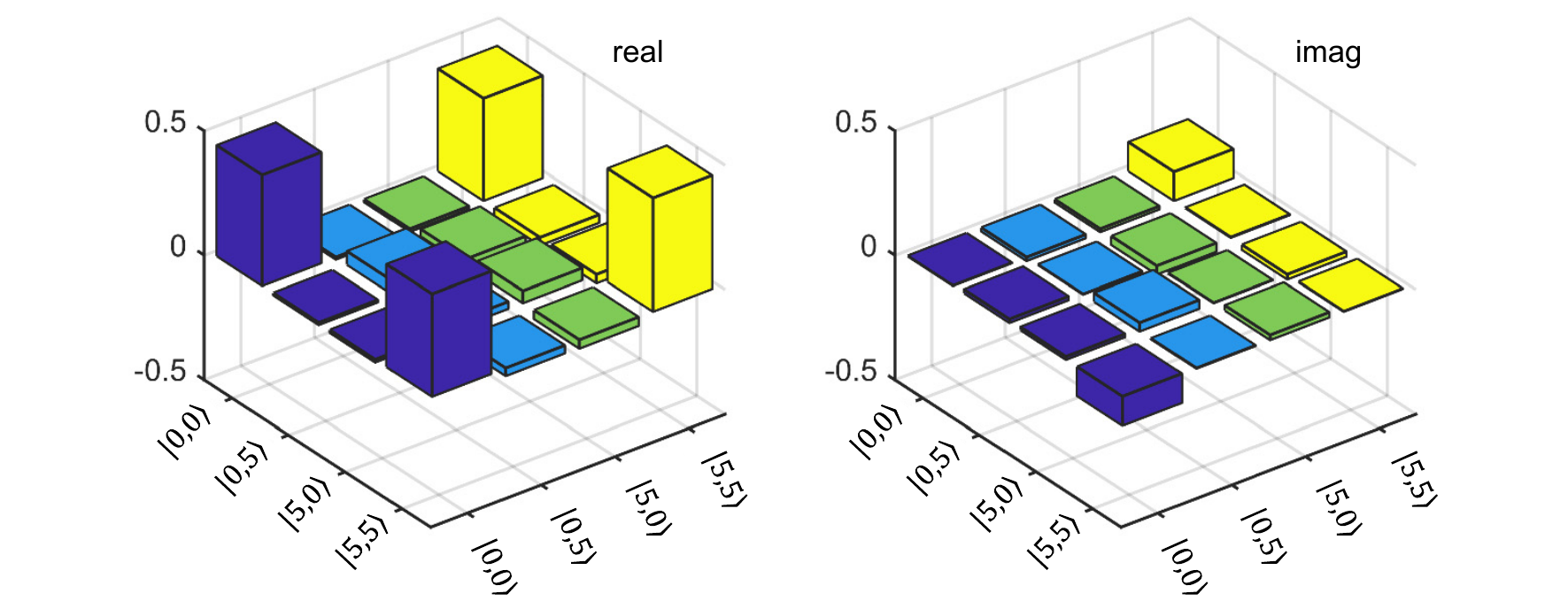}\\
\caption{\textbf{Density matrix of a 2-dimensional subspace after post-selection.} The density matrix is reconstructed by quantum state tomography. Here we choose a subspace spanned by $|K=0\rangle$ and $|K=5\rangle$ modes. The fidelity is $87.8\pm2.1\%$, with a relative phase of $17^{\circ}$ between the two modes.}
\end{figure*}

\section{Accidental coincidence}

The accidental coincidence is one of the major sources of noise in our experiment. The recorded photon coincidence can be written as $C_{SI}=(\eta_rP_S+P_SP_I)N$,
where $\eta_r$ is the retrieval efficiency of the spin wave, $P_S$ ($P_I$) denotes the probability to detect a signal (idler) photon, and $N$ is the total number of experiments. The second term is the random coincidence; in principle it will vanish if both $P_S$ and $P_I$ decrease simultaneously, since $\eta_r$ remains constant. However, at low excitation rate the photon counting will be dominated by the background noise; also the retrieval efficiency can drop when $P_S$ is low \cite{Laurat:06}.

In our experiment, the signal recording probability is maintained at $P_S\approx0.6\%$, and the experimental repetition rate is about $16000\,$s$^{-1}$. The accidental coincidence is subtracted as $C_{SI}^{\prime}=C_{SI}-C_{S}C_{I}/N$,
where $C_S=P_S N$ and $C_I=P_I N$ are the collected signal and idler photon counts respectively. After this correction, the results for entanglement witness, entanglement of formation and Bell-type inequality can be modified correspondingly.

\section{Quantum state tomography in subspace}

When measuring the visibility, we find that there is a relative phase between two $K$-space modes. Here, we measure a two-dimensional subsystem of $K$-space spanned by $|K=0\rangle$ and $|K=5\rangle$. Figure~6 is the reconstructed density matrix of the subspace by quantum state tomography after post-selection (renormalizing the trace to 1) \cite{PhysRevA.64.052312}. The fidelity is $87.8\pm2.1\%$ between the measured density matrix and the ideal one $|\Psi\rangle=(|0,0\rangle+|5,5\rangle)/\sqrt{2}$. A relative phase of $17^{\circ}$ exists owing to the basis setting difference between W-AOD and R-AOD. As is mentioned in Appendix~B, to eliminate the unwanted tones in AODs due to the nonlinear effect, the amplitudes and phases of the RF signals are optimized for each AOD. The optimal amplitude and phase settings for W-AOD and R-AOD are different, which effectively causes a relative phase between two modes in the $K$-space. This is another limit to obtain high entanglement dimension.



\bibliography{reference}

\end{document}